\documentclass[prd,aps,twocolumn,showpacs,floats,floatfix]{revtex4}
\usepackage[dvips]{graphicx}
 
\begin{document}
 
\title{Nonlinear $r$-Modes in Neutron Stars: Instability of an 
unstable mode}
 
\author{Philip Gressman}
\author{Lap-Ming Lin\footnote{Corresponding author.}}
\author{Wai-Mo Suen\footnote{Also at Department of Physics, The
Chinese University of Hong Kong, Hong Kong.}}
\affiliation{McDonnell Center for the Space Sciences, 
Department of Physics, Washington University, St. Louis, 
Missouri 63130}

\author{N. Stergioulas}
\affiliation{Department of Physics, Aristotle University of 
Thessaloniki, Thessaloniki 54006, Greece}
 
\author{John L. Friedman}
\affiliation{Department of Physics, University of Wisconsin-Milwaukee, 
PO Box 413, Milwaukee, Wisconsin 53201}
 
\date{\today}
 
\begin{abstract} 
We study the dynamical evolution of a large amplitude
$r$-mode by numerical simulations. $R$-modes in neutron stars are
unstable growing modes, driven by gravitational radiation reaction. In
these simulations, $r$-modes of amplitude unity or above are 
destroyed by a
catastrophic decay: A large amplitude $r$-mode gradually leaks energy into
other fluid modes, which in turn act nonlinearly with the $r$-mode,
leading to the onset of the rapid decay. As a result the $r$-mode suddenly
breaks down into a differentially rotating configuration. The
catastrophic decay does not appear to be related to shock waves at the
star's surface. The limit it imposes on the $r$-mode amplitude is
significantly smaller than that suggested by previous fully nonlinear
numerical simulations. 
\end{abstract}


\maketitle

%%%%%%%%%%%%%%%%%%%%%%%%%%%%%%%%%%%%%%%%%%%%%%%%%%%%% 
%\section{\bf Introduction} 
%%%%%%%%%%%%%%%%%%%%%%%%%%%%%%%%%%%%%%%%%%%%%%%%%%%%% 
%\paragraph*{Introduction.}
%%%%%%%%%%%%%%%%%%%%%%%%%%%%%%%%%%%%%%%%%%%%%%%%%%%%

The $r$-modes of rotating neutron stars are unstable growing modes
driven by gravitational radiation reaction \cite{Andersson:1,Friedman:1}.  
If the $l=m=2$ $r$-mode of a young, rapidly rotating star can grow
to an amplitude of order unity, the gravitational radiation it emits
would carry away most of the star's angular momentum and rotational
kinetic energy; and the radiation might be detectable by the Laser
Interferometric Gravitational Wave Observatory (LIGO II)
\cite{Owen} (for recent reviews see \cite{Andersson:2,Friedman:2}).
Even if its amplitude were smaller, the $r$-mode instability would limit
the periods of hot, young neutron stars (and, possibly, of old stars
spun up by accretion).  A number of mechanisms to damp the mode 
have been examined, including shear viscosity enhanced by crust-core
coupling and by nonstandard cooling \cite{crust}; bulk viscosity 
enhanced by a hyperon-rich core \cite{hyperon}; and energy loss 
to a magnetic field driven by differential rotation \cite{diff_rot}. 
But none of these definitively eliminates the instability.

The significance of the $r$-mode instability depends strongly on its
maximum possible amplitude.
In two recent Letters, Stergioulas and Font \cite{Nick:1} and  
Lindblom, Tohline, and Vallisneri \cite{LB:1} performed numerical  
simulations of nonlinear $r$-modes, both finding that  
large-amplitude nonlinear $r$-modes can exist for some long period of time. 
In addition, in \cite{LB:1,LB:2}, Lindblom {\it et al.} carried out 
numerical simulations of the growth of the $r$-modes driven by the current  
quadrupole post-Newtonian radiation reaction force in Newtonian  
hydrodynamics.   
In order to achieve a significant growth of the $r$-mode amplitude in a  
reasonably short computational time, they artificially  
multiplied the radiation reaction force by a factor of 4500.   
This decreases the growth time of the $r$-mode from about 40 s to 10 ms.   
The (dimensionless) $r$-mode amplitude $\alpha$ grew to $\approx 3.3$ before
shock waves appeared on the surface of the star and the $r$-mode amplitude 
collapsed. Lindblom {\it et al.} suggested that the nonlinear saturation 
amplitude of the $r$-modes may be set by dissipation of energy in the 
production of shock waves. 
Here, we show that a hydrodynamical effect will restrict  
the $r$-mode amplitude to a value significantly below that reported in 
\cite{LB:1,LB:2}. 
 
We note that, with the artificially large radiation reaction, the
results in \cite{LB:1,LB:2} assume that no hydrodynamic process takes
energy from the $r$-mode in a time scale between 10 ms and 40 s (the
artificial growth time and the actual physical growth time,
respectively).  In this paper we investigate the evolution of large
amplitude $r$-modes in these time scales (10 ms-40 s).  We find that
(i) a catastrophic decay of the mode, occuring within these time
scales, significantly reduces the amplitude to which an $r$-mode can
grow, and (ii) in a large amplitude $r$-mode this catastrophic decay
leads to a differentially rotating configuration.  (As in
Refs. \cite{LB:1,LB:2}, we assume a perfect fluid with no magnetic
fields.)
 
%%%%%%%%%%%%%%%%%%%%%%%%%%%%%%%%%%%%%%%%%%%%%%%%%%%%% 
%\section{\bf Description of The System } 
%%%%%%%%%%%%%%%%%%%%%%%%%%%%%%%%%%%%%%%%%%%%%%%%%%%%% 
%\paragraph*{Description of The System. }  
%%%%%%%%%%%%%%%%%%%%%%%%%%%%%%%%%%%%%%%%%%%%%%%%%%%%%%

We solve the Newtonian hydrodynamics equations for a non-viscous fluid  
flow in the presence of gravitational radiation reaction: 
\begin{equation} 
{\partial \rho\over \partial t}+\nabla\cdot\left(\rho \vec{v}\right)=0 \ , 
\label{eq:rho} 
\end{equation} 
\begin{equation} 
\rho\left({\partial \vec{v}\over \partial t}+ \vec{v}\cdot\nabla\vec{v}\right) 
=-\nabla P - \rho\nabla \Phi + \rho \vec{F}_{\rm GR}\ , 
\label{eq:Euler} 
\end{equation} 
where $\rho$ is the density, $P$ is the pressure, $\vec{v}$ is the velocity, 
$\vec{F}_{\rm GR}$ is the radiation reaction force per unit mass, 
and $\Phi$ is the Newtonian potential, satisfying  
\begin{equation} 
\nabla^2\Phi = 4 \pi G\rho \ . 
\label{eq:possion} 
\end{equation} 

A high resolution shock capturing scheme (Roe solver) is used to solve 
the hydrodynamic equations. In addition, 
as in \cite{Nick:1}, we applied the 3rd order piecewise parabolic 
method (PPM) \cite{Collela84} for the cell-reconstruction process
in order to simulate rapidly rotating stars
accurately for a large number of rotational periods.

As $r$-modes couple to the gravitational radiation mainly through  
the current multipoles $J_{lm}$, we assume that the contribution to  
the reaction force $\vec{F}_{\rm GR}$ comes solely from the dominant 
current multipole $J_{22}$, an approximation used also in \cite{LB:1,LB:2}.    
The resulting expression for $\vec{F}_{\rm GR}$ is given by 
(see \cite{Luc:1,Rez,LB:1}) 
\begin{equation} 
F_{\rm GR}^x - iF_{\rm GR}^y = -\kappa i\left(x+iy\right) 
\left[ 3v^z J_{22}^{(5)} + zJ_{22}^{(6)} \right] \ , 
\end{equation} 
\begin{equation} 
F_{\rm GR}^z = -\kappa \ {\rm Im}\left\{\left(x+iy\right)^2\left[ 
3{v^x+iv^y\over x+iy}J_{22}^{(5)} + J_{22}^{(6)} \right]\right\} \ , 
\end{equation} 
where $J_{22}^{(n)}$ is the nth time derivative of $J_{22}$ and
$\kappa =\kappa_{0}\equiv 32\sqrt{\pi}G/(45\sqrt{5} c^7)$.
A technical difficulty in evaluating $\vec{F}_{\rm GR}$  
is that it depends on high-order time derivatives of $J_{22}$.  
To circumvent this problem, we assume that 
$J_{22}^{(n)}=(i\omega)^{n}J_{22}$,  
with the nonlinear $r$-mode frequency $\omega$ defined by  
$\omega = - |J_{22}^{(1)}| / |J_{22}|$, as in   
\cite{LB:1,LB:2}.  
Notice that $J_{22}^{(1)}$ can be expressed as an integral  
over the fluid variables and is thus calculated on each time slice 
\cite{LB:1,LB:2,Rez}.  
 
%%%%%%%%%%%%%%%%%%%%%%%%%%%%%%%%%%%%%%%%%%%%%%%%%%%%% 
%\section{\bf Setting Up a Large Amplitude $R$-mode} 
%%%%%%%%%%%%%%%%%%%%%%%%%%%%%%%%%%%%%%%%%%%%%%%%%%%%% 
%\paragraph*{Setting Up a Large Amplitude $R$-mode. }  
%%%%%%%%%%%%%%%%%%%%%%%%%%%%%%%%%%%%%%%%%%%%%%%%%%%%%%

To investigate the dynamical properties of a large amplitude $r$-mode 
numerically, one must first generate initial data.   
However, analytically we know only the expression of an $r$-mode in  
the linear regime, under the assumption of $\alpha \ll 1$ and $\Omega_0 
\rightarrow 0$. In particular, the $l=m=2$ $r$-mode perturbation (at 
the first order of $\Omega_0$) is given by \cite{LB:3} 
\begin{equation} 
\delta \vec{v}= \alpha_{0}R_0 \Omega_{0}\left(r\over R_0 \right)^2  
 \vec{Y}_{22}^{B} e^{i\omega t} \ , 
\label{eq:delta_v} 
\end{equation} 
where $\alpha_0$ is a dimensionless amplitude, $R_0$ and $\Omega_0$ 
are the radius and angular velocity of the unperturbed rotating star 
model, and $\vec{Y}_{22}^{B}$ is the magnetic-type vector spherical 
harmonic defined by 
$\vec{Y}_{lm}^{B}=[l(l+1)]^{-1/2} \vec r \times \nabla Y_{lm}$.  
Using Eq. (\ref{eq:delta_v}) with a 
large $\alpha_0$ introduces other modes in the initial 
data \cite{LB:2}; that is, the resulting configuration is not the same as 
an $r$-mode growing from small 
amplitude driven by gravitational radiation reaction. In the following,  
we will show how we obtain a large amplitude $r$-mode in our study. 
 
Beyond the linear regime, there is no unique mode decomposition, and hence no 
unique definition of an $r$-mode.  In this paper, by a large amplitude 
$r$-mode we mean a configuration resulting from the growth of an 
infinitesimal $r$-mode due to gravitational radiation reaction.  
We define the amplitude $\alpha$ of the nonlinear $l=m=2$ mode in terms
of its contribution to $J_{22}=\int{\rho r^2 \vec{v}\cdot 
\vec{Y}_{22}^{B*} d^3x}$:
\begin{eqnarray} 
&&\alpha(t) \equiv 
\left|{{1\over R^3}  \int{ \tilde{\alpha}(x) e^{i\phi (x)} d^3x }}\right| 
\ , \cr
&&\cr
&&\tilde{\alpha}(x) e^{i\phi (x)} \equiv
{8\pi R^4 \left(\rho r^2 \vec{v}\cdot \vec{Y}_{22}^{B*}\right)\over
\bar{\Omega}(t)\int{\rho r^4 d^3x} }  \ . 
\label{eq:alpha} 
\end{eqnarray} 
Here $R$ is the radius of the corresponding nonrotating star model 
and $\tilde{\alpha}(x)$ is the amplitude density.  
The phase factor $\phi(x)$ is defined so that $\tilde{\alpha}(x)$ is 
real. 
The definition of the amplitude is similar to that of \cite{LB:1}, except 
that we normalize $\alpha(t)$ by the average angular velocity of the 
star $\bar{\Omega}(t)$ instead of a fixed initial $\Omega_0$. 
 
\begin{figure} 
\includegraphics[width=5.0cm]{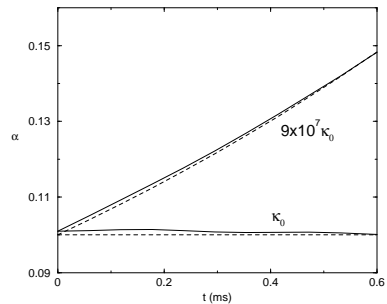} 
\caption{Evolution of $\alpha$ in a slowly rotating star with the  
correct ($\kappa_0$) and artificial ($9\times 10^7 \kappa_0$)
radiation reaction. The solid lines represent 
the numerical results ($257^3$ resolution), while the dashed lines  
are the predictions from linear theory.}  
\label{fig:257_slowtest} 
\end{figure} 
 
In Fig. \ref{fig:257_slowtest}, we show the time evolution of  
$\alpha$ for two cases in a slowly rotating star with  
$T=4.42\ {\rm ms}$.   
In the first case $\kappa$ is set to the correct Post-Newtonian  
value $\kappa_0$ (represented by the solid line labeled ``$\kappa_0$'').   
The evolution begins with a small (linear) $r$-mode perturbation  
given by Eq. (\ref{eq:delta_v}) with $\alpha_0=0.1$.   
The simulation is carried out up to $t=0.6\ {\rm ms}$. 
To compare, we plotted as dashed line the evolution of $\alpha$ 
as predicted by the linear theory \cite{LB:3}:   
$\alpha=\alpha_0 e^{t/ \tau_{\rm GR} }$, 
where the gravitational radiation time scale is given by 
\begin{equation} 
{1\over \tau_{\rm GR} } = {G\Omega_{0}^6\over 2 c^7} 
\left({256\over 405}\right)^2 \left({\kappa\over\kappa_0}\right) 
\int{\rho r^4 d^3x} \ . 
\end{equation} 
We note that this formula is correct only to linear order in $\alpha$ and 
in $\Omega$, and is thus accurate only for (i) $\alpha<<1$ and 
(ii) $\Omega_0/\sqrt{G\bar\rho} <<1$. While assumption 
(i) is reasonably good, (ii) is actually not accurate for the model used:  
rotation period 4.42 ms, corresponding to  
$\Omega / \Omega_{\rm max} \approx0.25$ ($\Omega_{\rm max}\approx  
2\sqrt{\pi G\bar{\rho} }/3$ is the Kepler limit,  
where $\bar{\rho}$ represents the average density of the star). 
Nevertheless, the two lines nearly coincide. 

With radiation reaction coefficient $\kappa_0$, an evolution time   
$t\approx 3\times 10^{8}\ {\rm ms}$ is needed to reach $\alpha=1$.    
For a $129^3$ grid-point simulation, this would
take $O(10^{11})$ time steps, requiring a clearly impractical 
$O(10^8)$ hours on a 128 CPU Origin 2000 (MIPS R12000).

To arrive at a large amplitude $r$-mode we use an artificially large $\kappa$  
as in \cite{LB:1,LB:2}.  In Fig. \ref{fig:257_slowtest}, the solid line  
labeled ``$9\times 10^7 \kappa_0$'' represents the evolution of $\alpha$  
with $\kappa = 9\times 10^7 \kappa_0$.   
For comparison, we also plot the evolution of $\alpha$ as predicted by  
the linear theory with this $\kappa$ as the dashed line.  The slight 
offset between the dashed and solid lines at the initial time is due to 
the fact that $\alpha(t=0)$ as calculated from Eq. (\ref{eq:alpha})  
equals to $\alpha_0$ as defined in Eq. (\ref{eq:delta_v}) only in the  
limit $\alpha_0 \ll 1$ and $\Omega_0 \rightarrow 0$. 
 
\begin{figure} 
   \centering 
   \includegraphics[width=3.8cm]{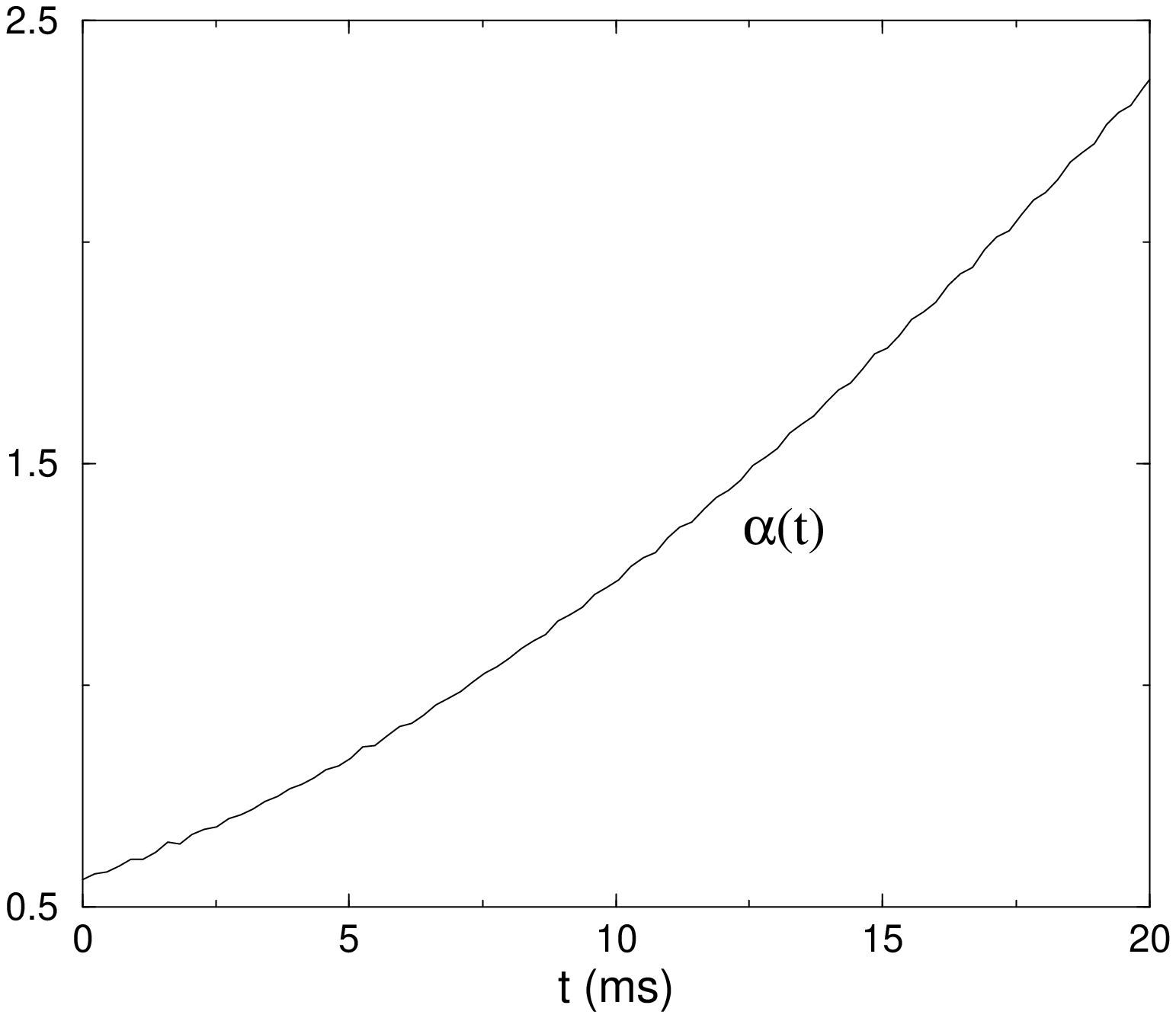}% 
   \includegraphics[width=4.2cm]{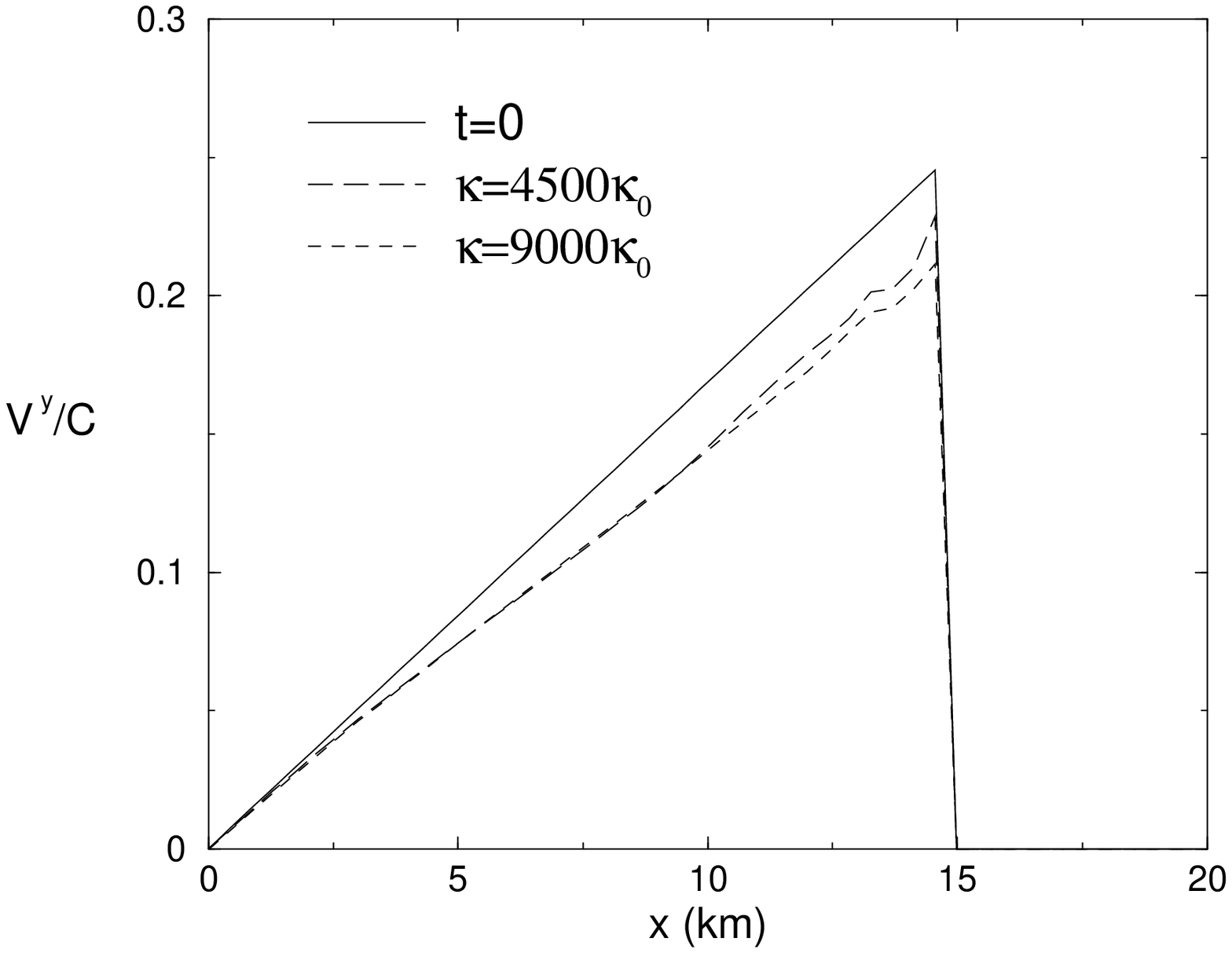}	 
   \caption{Left: Growth of the $r$-mode amplitude $\alpha$ during artificial 
            ``pumping'' with $\kappa=4500\kappa_0$. 
            Right: The velocity profile $v^y$ along the $x$ axis at $t=0$
             and at the point when $\alpha=2.0$.}
\label{fig:4500_2.2} 
\end{figure}

In Fig. \ref{fig:4500_2.2} (left), we show the growth of an $r$-mode  
(with $\alpha_0=0.5$) to a large amplitude with an artificial $\kappa$  
of $4500 \kappa_0$ in our fast rotating star model:  
The star has a mass of $M=1.64M_{\odot}$ with a polytropic equation of 
state $P=k\rho^2$.  The equatorial radius $R_{e}=14.5\ {\rm km}$.  
The ratio of the polar to equatorial radii is 0.76. The rotation period 
is $T=1.24$ ms.  
This model is used for the rest of the simulations discussed in this paper.   
Unless otherwise 
noted, we use $129^3$ Cartesian grid points with $\Delta x=\Delta 
y=\Delta z=0.42\ {\rm km}$.  
 
In Fig. \ref{fig:4500_2.2} (left), the amplitude 
$\alpha$ rises from 0.5 to 2.2 in 19 ms.  
An indicator of the accuracy of the simulation is that the total mass
of the system is constant to 0.08\% by 20 ms in our $129^3$ runs.
Also, the actual numerical evolution of the total angular momentum 
$J=|\int{\rho \vec{r} \times \vec{v} d^3x}|$ agrees with the 
theoretical prediction [see Eq. (11) of Ref. \cite{LB:1}] to about 
1\%.
 
%%%%%%%%%%%%%%%%%%%%%%%%%%%%%%%%%%%% 
 
While the discussion above shows the accuracy of our numerical treatment, 
one still must ask whether the large amplitude $r$-mode obtained 
with the large artificial pumping is physical or not,
in the sense that whether the rapid pumping excites modes that 
would not be excited with $\kappa = \kappa_0$.  
We compare the large amplitude $r$-modes obtained with different 
pumping rates and conclude that the resulting fluid flow pattern
does not depend sensitively on the pump rate (as long as the pump 
rate is large enough so that the large amplitude $r$-mode can be
arrived at).
In Fig. \ref{fig:4500_2.2} (right), we 
show the rotational velocity profile $v^y$ along the $x$ axis for 
$\kappa = 9000,\ 4500 \kappa_0$ at the point when $\alpha =2.0$, 
starting with the same initial model.  For comparison, we also plot 
the initial profile in the same figure.  We see that the two lines, 
$\kappa=9000,\ 4500\kappa_0$, agree with each other to better than 
3\%, with smaller discrepancy away from the surface.    
In the rest of this paper, we will take the 
``pumped up'' configurations given in Fig. \ref{fig:4500_2.2} (left) 
as the initial state in our investigation of the hydrodynamical behavior of 
the large amplitude $r$-mode.  To the extent that different pumping
rates are not affecting the initial state we use, the artificial
pumping is not affecting the results we report below.

%%%%%%%%%%%%%%%%%%%%%%%%%%%%%%%%%%%%%%%%%%%%%%%%%%%%%%%%%%%%%%%%% 
%\section{Catastrophic nonlinear decay of a large amplitude $R$-mode} 
%%%%%%%%%%%%%%%%%%%%%%%%%%%%%%%%%%%%%%%%%%%%%%%%%%%%%%%%%%%%%%%%%% 
%\paragraph*{Catastrophic nonlinear decay of a large amplitude $R$-mode.} 
%%%%%%%%%%%%%%%%%%%%%%%%%%%%%%%%%%%%%%%%%%%%%%%%%%%%%%%%%%%%%%%%%%%% 

Here we focus on one question: What is the fate of a large-amplitude
$r$-mode in a rapidly rotating neutron star for a sufficiently long 
time evolution, modeled as the 1.24 ms polytrope described above.  
That is, what evolution is implied by Eqs. (\ref{eq:rho})-(\ref{eq:possion}), 
with the correct amount of radiation reaction?
\begin{figure} 
   \centering 
   \includegraphics[width=4.0cm]{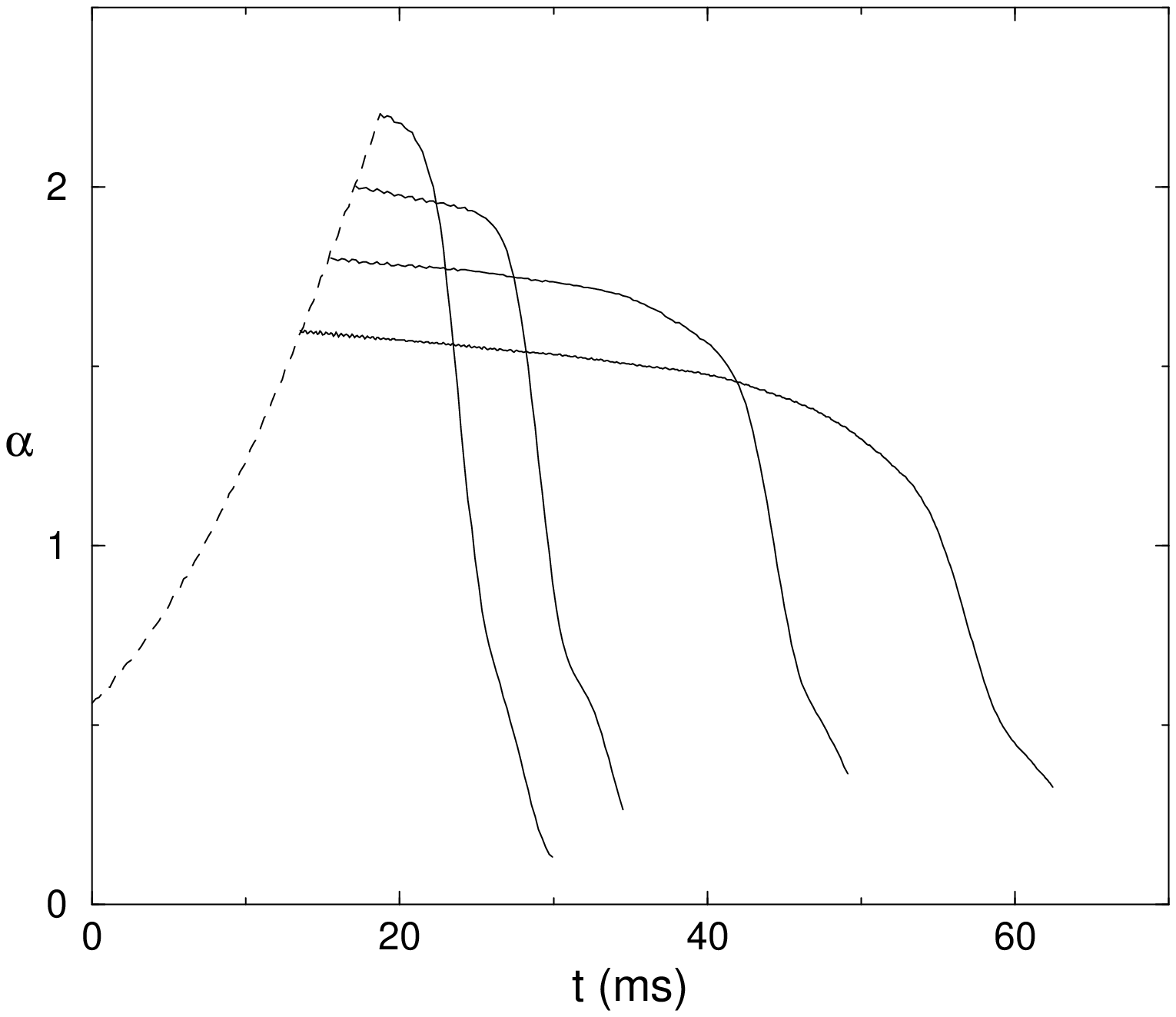}% 
   \includegraphics[width=4.0cm]{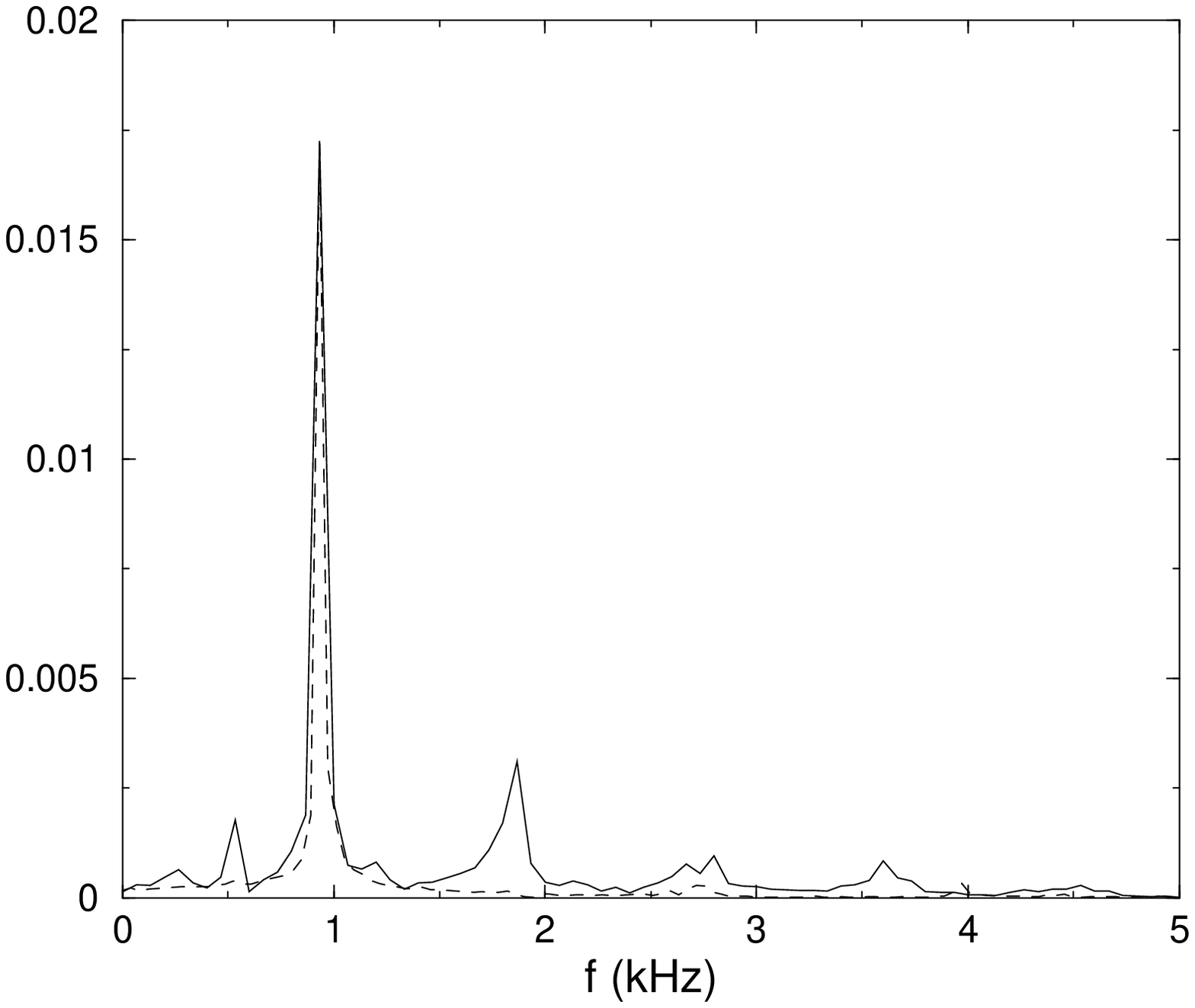} 
    \caption{Left: Evolution of the ``pumped up'' $\alpha$  
             at different values with $\kappa=\kappa_0$. 
             Right: Fourier spectra of $v^z$ along the $x$ axis (at  
		    $x=6$ km) at two time slots in the evolution 
	 	    starting out with $\alpha=1.6$. The early (later) time  
		    slot is denoted by the dashed (solid) line.} 
\label{fig:4500_alpha} 
\end{figure} 
 
In Fig. \ref{fig:4500_alpha} (left) we show the evolution of $\alpha$ vs  
time for various large amplitude $r$-modes starting off with  
$\alpha=2.2,\ 2.0,\ 1.8,\ 1.6$.  We see  
that the mode amplitudes start off slowly decaying, leaking energy to other  
modes.  
The decay rate is small, until a certain time.
We plot in Fig. \ref{fig:4500_alpha} (right) the Fourier transform  
of the velocity component $v^z$ along the $x$ axis 
at a typical point inside the star at two different time slots in the  
evolution of the $\alpha=1.6$ case [the lowest line in  
Fig. \ref{fig:4500_alpha} (left)]. 
The dashed line is the profile at the beginning of the evolution 
(13 ms-35 ms), we see that there is only one large peak at the  
$r$-mode frequency (0.93 kHz). 
This is compared to the solid line representing the spectrum at a later time  
slot (35 ms - 50 ms). We see that various smaller peaks appear in the  
spectrum, especially the one at twice the $r$-mode frequency (1.86 kHz).  
Spectra at different points inside the star give similar  
structure, with those in the core region showing more peaks at
different frequencies.
 
The most interesting feature of Fig. \ref{fig:4500_alpha} (left) is 
that after some slow leaking of energy into other fluid modes, the $r$-mode 
amplitude drops catastrophically to a value much smaller than 1. 
This abrupt drop occurs through nonlinear couplings with other fluid 
modes: In the slow leaking phase, these other fluid modes
are growing linearly until a certain unstable point.
The time it takes to reach the unstable point depends sensitively on the  
$r$-mode amplitude.  It shortens from approximately 45 ms to 8 ms when the  
initial value of $\alpha$ changes from 1.6 to 2.2.

\begin{figure} 
\includegraphics[width=4.8cm]{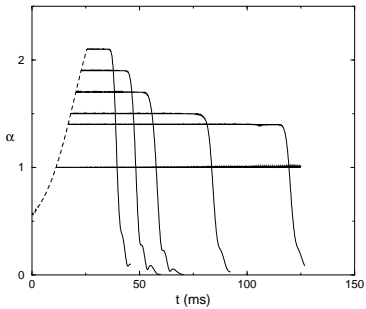} 
\caption{Evolution of $\alpha$ ($65^3$ resolution) with artificial pumping of 
$\kappa = 4500 \kappa_0$ whenever $\alpha$ drops below its initial value.} 
\label{fig:65_alpha} 
\end{figure}

To further investigate this catastrophic decay we perform a set of
numerical experiments in which we pump energy into the $r$-mode by
turning on the artificial radiation reaction force with coefficient
$\kappa =4500\kappa_0$ whenever its amplitude drops down below its
initial value.  The resulting evolution of $\alpha$ vs time is given
in Fig. \ref{fig:65_alpha}.  The evolution tracks of $\alpha$ starting
off with values $\alpha=2.1,\ 1.9,\ 1.7,\ 1.5,\ 1.4,\ 1.0$ are given.
With the large artificial pumping $\alpha$ remains constant despite
energy leaking to other modes.  In all cases (except $\alpha=1.0$
where the simulation is not evolved long enough), however, the
hydrodynamical nonlinear interaction eventually overwhelms the
artificial pumping, and the $r$-mode amplitude falls catastrophically.

\begin{figure}[b] 
   \centering 
   \includegraphics[width=4.0cm]{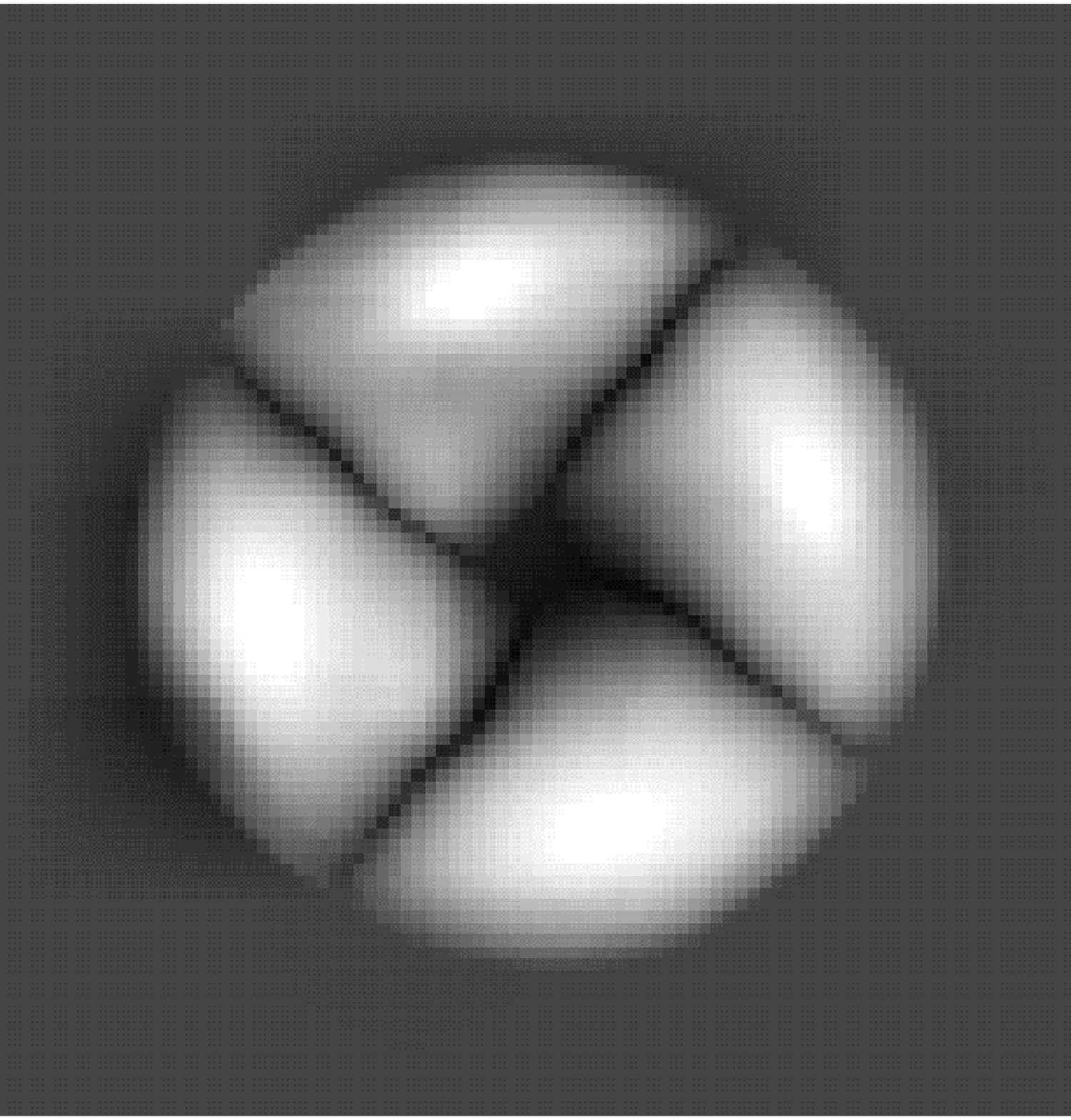}% 
   \includegraphics[width=4.0cm]{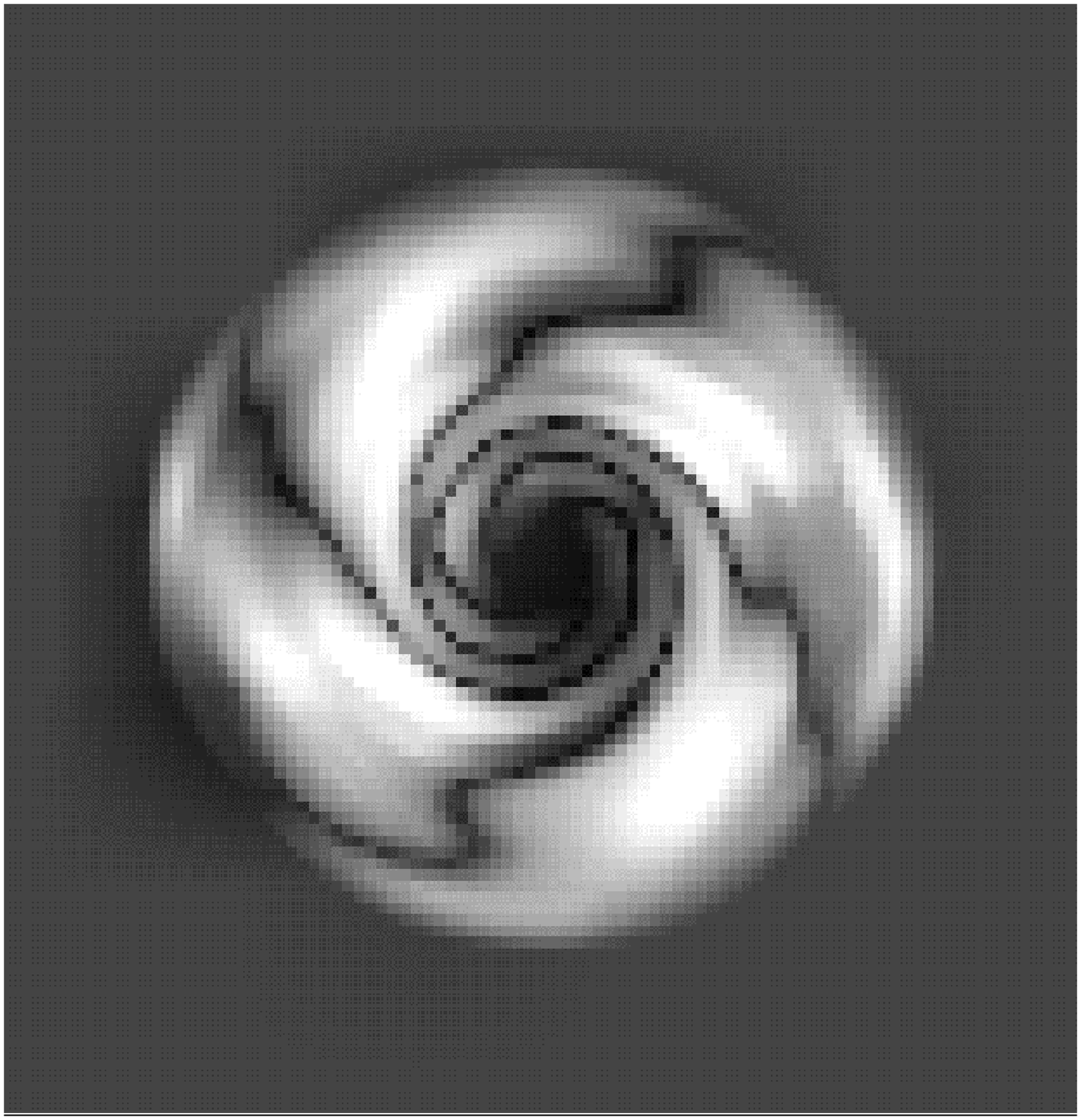} 
   \caption{The amplitude density $\tilde{\alpha}(x)$ on the equatorial 
            plane before (left) and after (right) the breakdown for the  
            case where the initial ``pumped up'' $\alpha$ is 2.0 in  
            Fig. \ref{fig:4500_alpha}.} 
\label{fig:2.0alpha_den} 
\end{figure}

In Fig. \ref{fig:2.0alpha_den} we compare the distribution of the amplitude  
density $\tilde{\alpha}(x)$ as defined in Eq. (\ref{eq:alpha})  
on the equatorial plane before (left) and after (right)  
the breakdown respectively for the case where the initial  
$\alpha$ is 2.0 (the $\alpha =2.0$ line in Fig. \ref{fig:4500_alpha}).  
In the figure, the brighter region represents higher amplitude density.  
During the catastrophic decay, the $r$-mode pattern changes  
rapidly from a 4-fold ``regular'' shape (left) to a whirlpool-like  
spiral (right).  
We also see in our simulations that strong differential rotation 
is developed during the breakdown, a potentially important fact regarding  
whether subsequent re-growth of the $r$-mode is possible or not  
\cite{Karino}. 

\begin{figure}
\includegraphics[width=4.5cm]{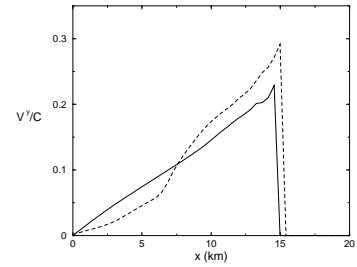} 
\caption{The rotational velocity profile $v^y$ along the $x$ axis for  
the case where the initial ``pumped up'' $\alpha$ is 2.0.  
The solid line is the initial profile at $\alpha=2.0$,  
while the dashed line is the profile after the breakdown.} 
\label{fig:vy_diffrot} 
\end{figure}

\begin{figure}[b] 
   \centering 
   \includegraphics[width=3.9cm]{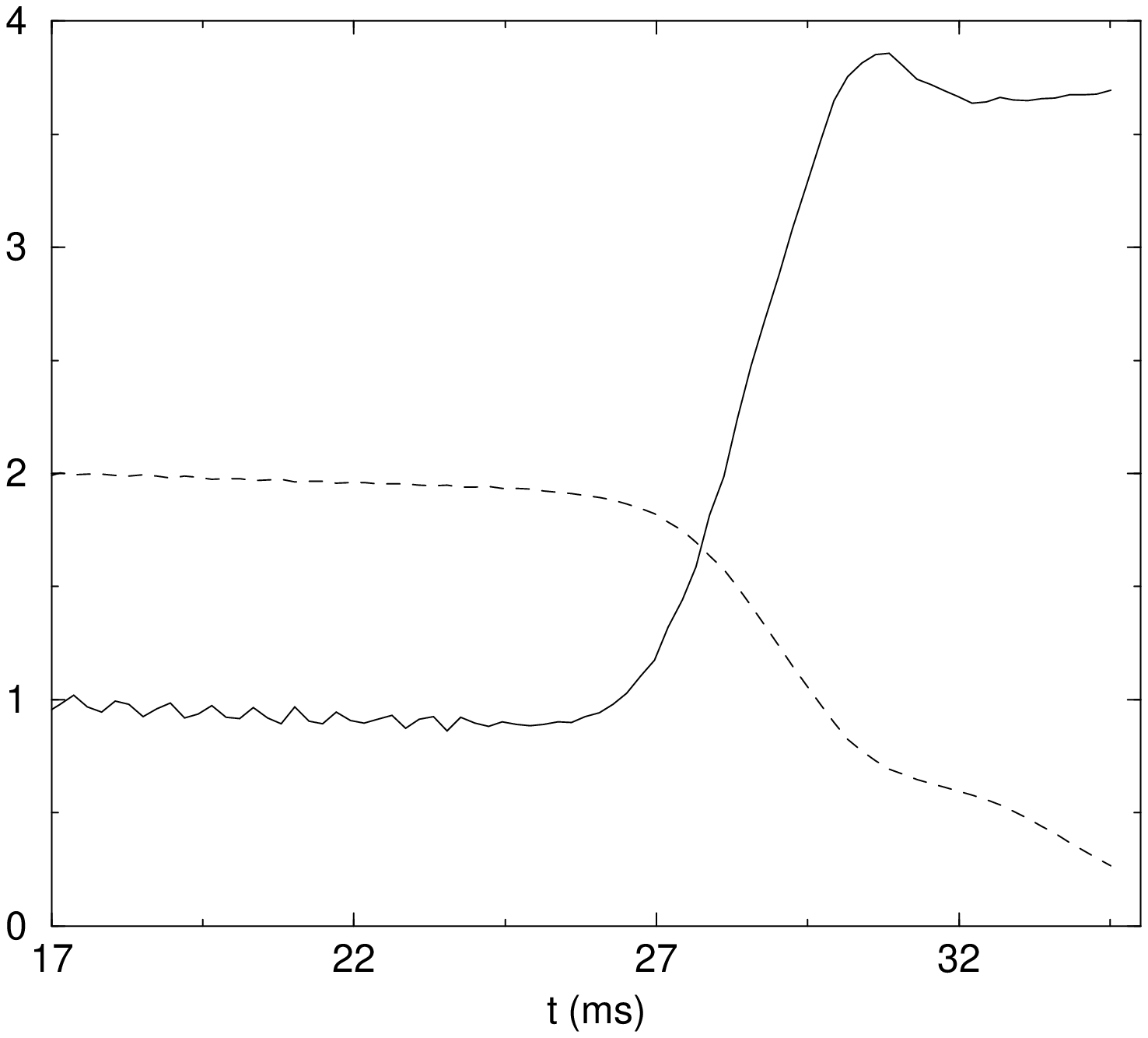}% 
   \includegraphics[width=4.1cm]{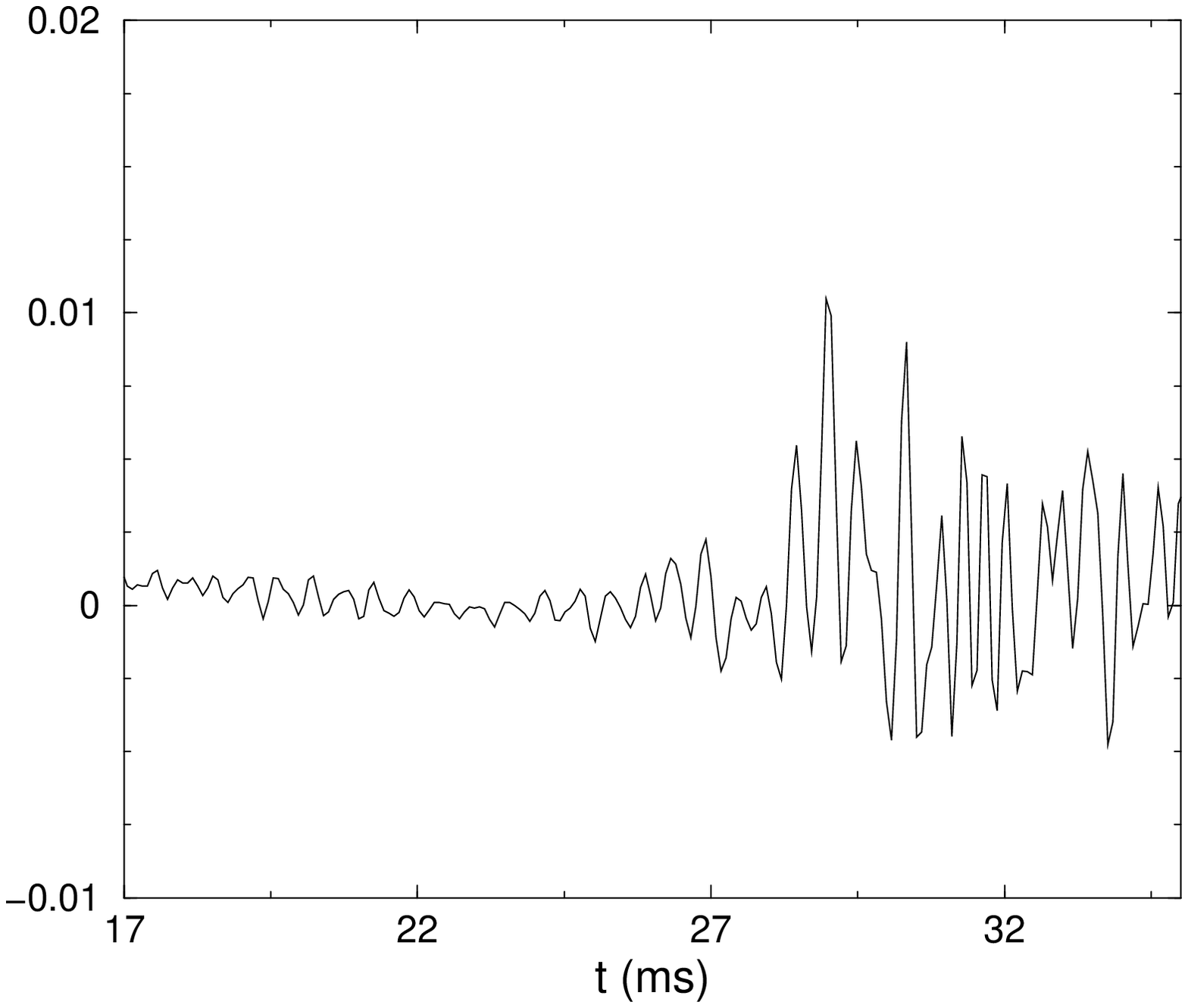} 
   \caption{Left: Evolution of $\alpha$ (dashed) and $I$ (solid) for the  
            case where the initial ``pumped up'' $\alpha$ is 2.0. 
            Note that $I$ has been rescaled for comparison.  
            Right: Evolution of $Q_{xy}$ (in $G=c=M_{\odot}=1$ units) 
            for the same case.}  
\label{fig:diff_rot} 
\end{figure}

In Fig. \ref{fig:vy_diffrot} we plot the rotational velocity profile  
$v^y$ along the $x$ axis for the case where the initial $\alpha$ is 2.0.  
The solid line is the initial profile, while the dashed line is the profile  
after the decay. 
To further quantify the amount of differential rotation, we define the  
kinetic energy associated to differential rotation by  
$I= {1\over 2}\int{ \rho \left( v_{\phi} - \bar{v}_{\phi}\right)^2 
d^3x }$, 
where $\bar{v}_{\phi}= \bar{\Omega}\sqrt{x^2+y^2}$ with $\bar{\Omega}$ 
being the average angular velocity of the star.  
We plot the evolution of $\alpha$ and $I$ together in  
Fig. \ref{fig:diff_rot} (left).  
It is seen that the amount of differential rotation ($I$) rises rapidly  
during the breakdown of $\alpha$.  
We also see in our simulations that the star has a relatively large 
amplitude pulsation during the breakdown.  
Fig. \ref{fig:diff_rot} (right) shows the quadrupole-moment 
component $Q_{xy}$ against time for the same case as in  
Fig. \ref{fig:diff_rot} (left).  
We see that $Q_{xy}$ is basically zero until the breakdown and it then 
oscillates rapidly afterward. However, based on an order-of-magnitude  
estimation, the gravitational radiation amplitude due to the changing  
quadrupole moment is only about 1\% of that due to the $r$-mode.

In contrast to the study of Refs. \cite{LB:1,LB:2}, 
we do not see evidence that  
this catastrophic decay is due to the generation of shock waves on the  
surface of the star.   
  
%%%%%%%%%%%%%%%%%%%%%%%%%%%%%%%%%%%%%%%%%%%%%%%%%%%%%%%%%%%%% 
%\section{Discussions and Conclusion} 
%%%%%%%%%%%%%%%%%%%%%%%%%%%%%%%%%%%%%%%%%%%%%%%% 
%%%%%%%%%%%%%%%%%%%%%%%%%%%%%%%%%%%%%%%%%%%%%%%% 
%\paragraph*{Discussions and Conclusion.} 
%%%%%%%%%%%%%%%%%%%%%%%%%%%%%%%%%%%%%%%%%%%%%%%% 
 
We found in this paper that a large amplitude $r$-mode will lose energy
to other fluid modes whose growth in turn trigger a catastrophic decay
of the $r$-mode.
For an $r$-mode with an amplitude $\alpha$ of order one, the onset 
of catastrophic decay requires a time much shorter than the growth time of 
the $r$-mode due to radiation reaction; and the decay thus limits the 
$r$-mode amplitude to a value less than that found by \cite{LB:1,LB:2}.   
Further work is in progress to determine the nature of this
catastrophic decay, whether it is related to any known hydrodynamical
instability of nonlinear flow, and how large an $r$-mode amplitude  
can be with this effect taken into account.   
 
{\it Note added.} Towards the end of the preparation of this paper we learned
the results of Arras {\it et al.} \cite{Arras}.  
We note the following differences between the hydrodynamical 
phenomenon studied in this paper and that studied by them:
The work of Arras {\it et al.} points to a slow leakage of the $r$-mode energy
into some short wavelength oscillation modes, leading to an equilibrium
distribution of mode amplitudes.  This in turn, through viscosity dissipation,
limits the $r$-mode amplitude to a small value.  In this paper we find
a sudden and complete breakdown of the $r$-mode that operates
independent of viscosity.  Further numerical
investigation will be carried out to investigate the interactions of
$r$-modes with other oscillation modes.  Such investigation is beyond
the resolution power of our present simulations.  Note that, if the
conclusions of \cite{Arras} are correct, there may be no astrophysical
situation in which $r$-modes grow to amplitudes large enough to exhibit
the sudden decay seen in our simulations.
 
%%%%%%%%%%%%%%%%%%%%%%%%%%%%%%%%%%%%%%%%%%%%%%%%%% 
%\paragraph*{Acknowledgments.}
%%%%%%%%%%%%%%%%%%%%%%%%%%%%%%%%%%%%%%%%%%%%%%%%%%%

We thank N. Andersson, G. Comer, E. Evans, S. Iyer, L. Lindblom, 
M. Miller, Y. Mino, B. Owen and C. Will for useful discussions.  
The simulations in this paper made use
of the following code components: Newton\_evolve (Newtonian gravity and
evolution) by P. Gressman, PPM (PPM reconstruction) by T. Font,
Newton\_anal (modules for various Newtonian analyses) by L.-M. Lin, and
the Cactus computational toolkit by T. Goodale {\it et al.}  
The research is supported by NSF Phy 00-71044, 00-96522, 99-79985
(KDI Astrophysics Simulation Collaboratory Project), NRAC MCS93S025,
the EU Programme ``Improving the Human Research Potential and the
Socio-Economic Knowledge Base'' (Research Training Network Contract
HPRN-CT-2000-00137), KBN-5P03D01721, and support from the NASA AMES NAS.
     
%%%%%%%%%%%%%%%%%%%%%%%%%%%%%%%%%%%%%% 
%% reference 
%%%%%%%%%%%%%%%%%%%%%%%%%%%%%%%%%%%%%% 
\bibliographystyle{prsty}

\begin{thebibliography}{1}
 
\bibitem{Andersson:1}N. Andersson, Astrophys. J. {\bf 502}, 
708 (1998). 
 
\bibitem{Friedman:1}J. L. Friedman and S. M. Morsink, Astrophys. J. 
{\bf 502}, 714 (1998). 
 
\bibitem{Owen}B. J. Owen {\it et al.}, Phys. Rev. D {\bf 58}, 084020 (1998). 
 
\bibitem{Andersson:2}N. Andersson and K. D. Kokkotas, 
Int. J. Mod. Phys. D {\bf 10}, 381 (2001). 
 
\bibitem{Friedman:2}J. L. Friedman and K. H. Lockitch, 
gr-qc/0102114. 
 
\bibitem{crust}N. Andersson, D. I. Jones, K. D. Kokkotas, and N. Stergioulas,
 Astrophys. J. {\bf 534}, L75 (2000);
 M. Rieutord, astro-ph/0003171;
 L. Lindblom, B. J. Owen and G. Ushomirsky, Phys. Rev. D {\bf 62},
 084030 (2000);
 Y. Levin and G. Ushomirsky, Mon. Not. R. Astron. Soc. {\bf 324}, 917 (2001); 
 Y. Wu, C. D. Matzner, and P. Arras, Astrophys. J. {\bf 549}, 1011 (2001);
 G. Mendell, Phys. Rev. D {\bf 64}, 044009 (2001).
 
\bibitem{hyperon} P. B. Jones, Phys. Rev. Lett. {\bf 86}, 1384 (2001); 
 P. B. Jones, Phys. Rev. D {\bf 64}, 084003 (2001); 
 L. Lindblom and B. J. Owen, Phys. Rev. D {\bf 65}, 063006 (2002); 
 P. Haensel, K. P. Levenfish, and D. G. Yakovlev, astro-ph/0110575.

\bibitem{diff_rot}H. C. Spruit, Astron. Astrophys. {\bf 341}, L1 (1999); 
L. Rezzolla, F. K. Lamb, and S. L. Shapiro, Astrophys. J. {\bf 531}, 
L139 (2000); L. Rezzolla, F. K. Lamb, D. Markovic,and S. L. Shapiro,
Phys. Rev. D {\bf 64}, 104013 and 104014 (2001).

\bibitem{Nick:1}N. Stergioulas and J. A. Font, Phys. Rev. Lett. 
{\bf 86}, 1148 (2001). 
 
\bibitem{LB:1}L. Lindblom, J. E. Tohline, and M. Vallisneri, Phys. Rev. 
Lett. {\bf 86}, 1152 (2001). 
 
\bibitem{LB:2}L. Lindblom, J. E. Tohline, and M. Vallisneri,
Phys. Rev. D {\bf 65}, 084039 (2002).
 
\bibitem{Collela84}
P. Collela and P. R. Woodward, J. Comput. Phys. {\bf 54}, 174 (1984).
 
\bibitem{Luc:1}L. Blanchet, Phys. Rev. D {\bf 55}, 714 (1997). 
 
\bibitem{Rez}L. Rezzolla {\it et al.}, Astrophys. J. {\bf 525}, 935 (1999). 
 
\bibitem{LB:3}L. Lindblom, B. J. Owen, and S. M. Morsink, Phys. Rev. Lett. 
{\bf 80}, 4843 (1998). 
 

\bibitem{Karino}S. Karino, S. Yoshida, and Y. Eriguchi, Phys. Rev. D
{\bf 64}, 024003 (2001).
 
\bibitem{Arras}P. Arras {\it et al.}, astro-ph/0202345. 
 
 
\end{thebibliography}

\end{document}